\documentclass{article}

\usepackage[margin=1in]{geometry} 
\usepackage{amsmath,amsthm,amssymb}
\usepackage{textcomp}
\usepackage[dvips,final, pdftex]{graphicx}         		
\usepackage[greek, english]{babel}                			
\usepackage[hang,bf]{caption}              			
\usepackage{rotating}                     		 		
\usepackage{subfigure}
\usepackage{units}
\usepackage{fancyhdr}
\usepackage{lscape}
\usepackage{amsfonts}
\usepackage{float}
\usepackage{graphicx}
\usepackage{longtable}
\usepackage{fancybox}
\usepackage[hidelinks]{hyperref}
\usepackage{here}
\usepackage{setspace} 
\usepackage{units} 
\usepackage{pdfpages}
\usepackage{lscape} 		
\usepackage{pdflscape}

\usepackage{upgreek}

\usepackage{listings}

\begin{document}

\title{Closed-form solutions for tilted three-part piecewise-quadratic half-plane contacts} 
\author{H. Andresen$^{\,\text{a,}}$\footnote{Corresponding author: \textit{Tel}.: +44 1865 273811; \newline \indent \indent \textit{E-mail address}: hendrik.andresen@eng.ox.ac.uk (H. Andresen).}$\,\,$, D.A. Hills$^{\,\text{a}}$, J.V\'azquez$^{\,\text{b}}$\\ \\
\small{$^{\text{a}}$ Department of Engineering Science, University of Oxford, Parks Road, OX1 3PJ Oxford, United Kingdom} \\
$\,$\small{$^{\text{b}}$ Departmento de Ingenier\'ia Me\'canica y Fabricaci\'on, Universidad de Sevilla, Camino de los Descubrimientos,} \\ 
\small{41092 Sevilla, Spain}}

\date{}

\maketitle

\begin{center}
	\line(1,0){470}
\end{center}
\begin{abstract}
  A general half-plane contact problem in which the geometry is specified in a piecewise-quadratic sense over three segments is solved in closed form. This includes the effects of a moment applied sufficient to introduce separation of one segment and the application of a shearing force sufficient or insufficient to cause sliding. Extending existing solutions to asymmetrical problems is necessary in order to broaden our understanding of the behaviour of dovetail roots of gas turbine fan blades.
  In previous studies symmetrical contacts have often been used to represent a dovetail flank contact. In the asymmetrical case, the contact pressure may be considerably higher at one of the contact edges compared to the corresponding symmetrical case. Exploiting the generality provided with the solution presented in this study, several simpler indenter problems are investigated making use of an algebraic manipulator. The Mathematica code is made available for download.  \\
  
  \end{abstract}
\begin{center}
	\line(1,0){470}
\end{center}

\section{Introduction}

\begin{figure}[H]
		\centering
		\includegraphics[scale=0.35, trim=0 0 0 0, clip]{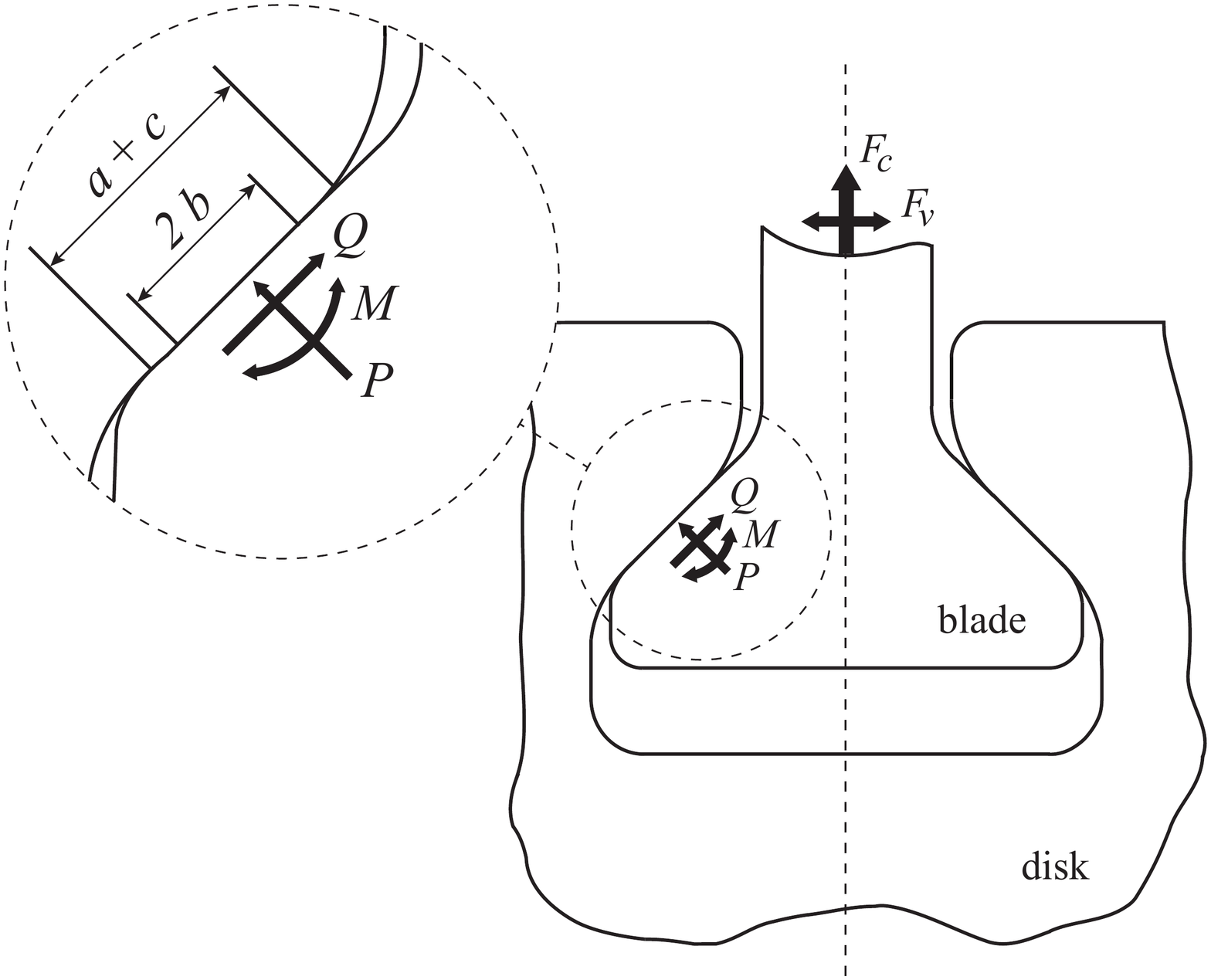}
		\caption{Illustration of a dovetail geometry subject to external loads}
		\label{fig:dovetail}
\end{figure}

\hspace{0.4cm} The reliability of a gas turbine engine is largely dependent on the structural integrity of its blades. Turbine and compressor blades are often connected to the disk by a dovetail joint as depicted in figure \ref{fig:dovetail}.  As the gas turbine rotates, the dovetail geometry experiences centrifugal, $F_c$, and vibrational, $F_v$, loading. The centrifugal force brings the blade and disk into contact. The two established contact flanks are subject to normal and shear forces, here indicated by $P$ and $Q$, respectively.  The curvature at each end of the contact will almost always be unequal, not only due to manufacturing uncertainties, but also as an intentional design. As a consequence, the tractions and contact interface will be asymmetrical. Furthermore, vibrational loads may introduce a moment, $M$, to the contact, which leads to additional asymmetry. 

The majority of solutions encountered in the literature assume symmetrical profiles and symmetrical indentation \cite{Vazquez_2009}, \cite{Ciavarella_1998_2}. This greatly simplifies the solution of the related contact problem but may be an over-simplification. Individual aspects of asymmetry in half-plane contacts have been studied before \cite{Davies_2010}, \cite{Sackfield_2005}.  However, the novelty here is to provide a set of versatile closed-form solutions for asymmetrical half-plane contacts, which are capable of considering  \textit{asymmetrical geometries} as well as the  \textit{application of a moment} during normal indentation. 
This scenario of asymmetrical indentation has not been addressed in the literature for the cases studied in this work.

\begin{figure}[H]
		\centering
		\includegraphics[scale=0.35, trim=100 1930 0 80, clip]{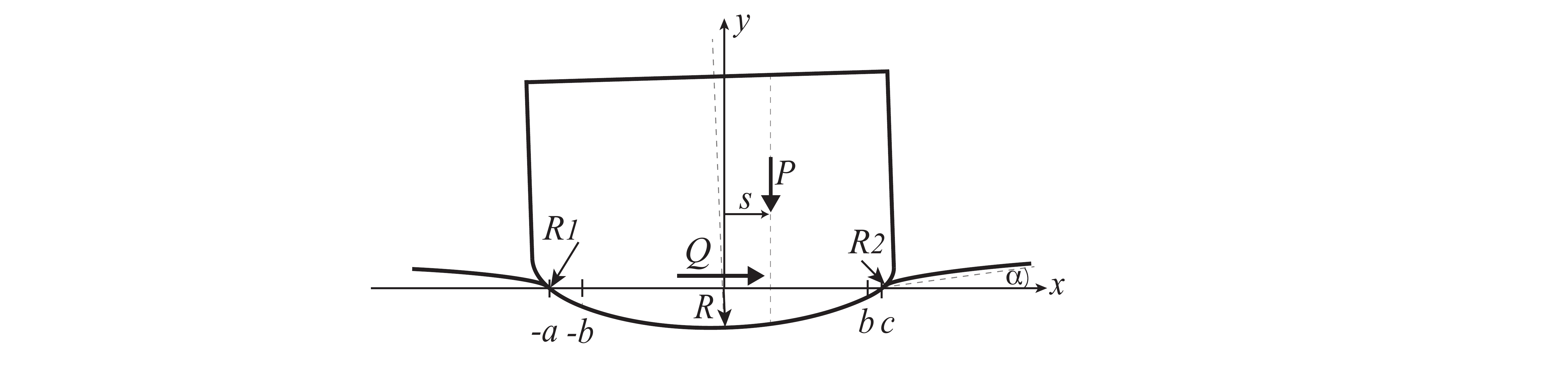}
		\caption{Illustration of the tilted punch comprised of three different rounded segments}
		\label{fig:illustration_generic_punch}
\end{figure}

 We chose to study, first, the most general case where the problem may be idealised by a punch with a compound curvature in contact with a half-plane as shown in figure \ref{fig:illustration_generic_punch}. The punch consists of a curved central portion of radius $R$ over the length $2b$, whereas the edges are rounded with radii $R_{1}$ and $R_{2}$ with a continuous gradient across the joins. This means the contact itself is incomplete in character and, given the outer radii are unequal, the geometry is asymmetrical. First, a normal load, $P$, is applied at a distance, $s$, from the centre line of the flat portion such that an either clockwise or counter-clockwise moment, $P s$, depending on the orientation of $s$, is introduced to the problem. The contact pressure, $p(x)$, the extent of the contact, $[-a, c]$, and the amount of tilt, $\alpha$, are to be quantified. The solution found is given in closed form. Subsequently, a monotonically increasing shearing force, $Q$, is applied in the plane of the contact interface. This will give rise to a partial slip pattern if the shear force is insufficient to cause sliding (that is, when $|Q| < fP$, where $f$ is the coefficient of friction). Using the Ciavarella-J\"ager theorem, the size and position of the stick zone can be quantified. 

\begin{figure}[H]
	\centering
	\includegraphics[scale=0.197, trim=0 510 0 600, clip]{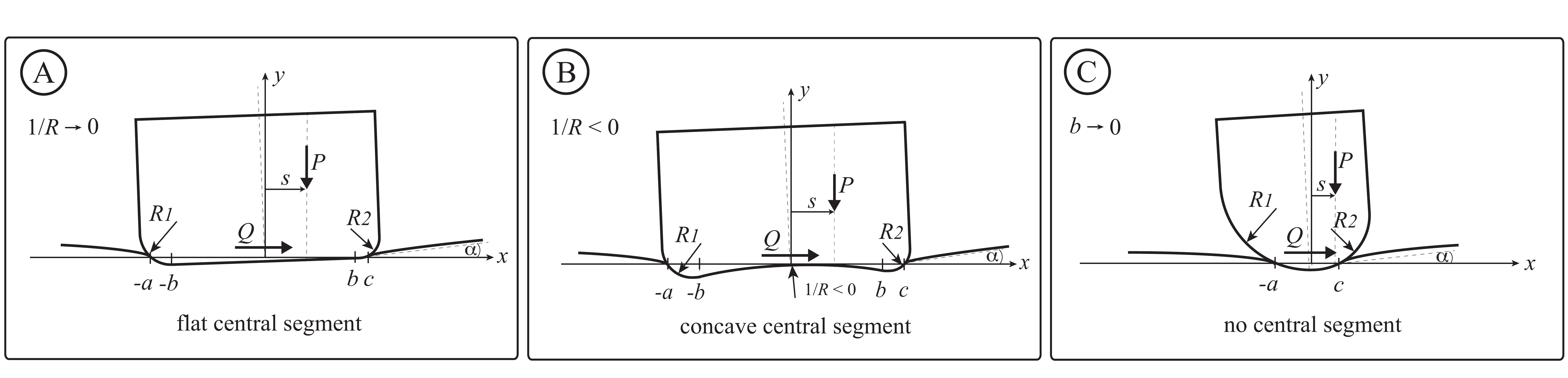}
	\caption{Derivations from the original problem statement}
	\label{fig:illustration_generic_punch_ABCD_1}
\end{figure}

The generality of the problem statement serves as an umbrella to address potentially several simpler indenter problems by making use of an algebraic manipulator. We chose to use the general solution found to study three specific cases. These include the tilted flat but rounded punch with unequal radii which is thought to be a suitable candidate for a dovetail flank contact representation. Figure \ref{fig:illustration_generic_punch_ABCD_1} depicts the three chosen derivations of the original problem statement. 

Modification `A' might be referred to as the tilted flat but rounded punch with unequal radii. Case `B' addresses the case when the central segment is concave, i.e. $1/R < 0$. Finally, derivation `C' is a tilted punch composed of two adjoined unequal half-cylinders. Although out of the three cases presented, modification `A' is thought to resemble a dovetail flank contact most accurately, the other cases might be of interest for certain engineering applications and add to the knowledge base of contact mechanics. It is noteworthy that despite the changes to the original problem definition, the solution derived for the corrective shear traction still applies in the same way as the normal pressure solution. Solutions to all cases, `A', `B', and `C', can be found in appendices \ref{appendix:caseA}, \ref{appendix:caseB}, and \ref{appendix:caseC}. For completeness, Mathematica code for each case is provided as a download. 

\section{Formulation}

\hspace{0.4cm}The three-part punch is formulated as a linear-elastic body within half-plane theory indenting a linear-elastic half-plane. However, the solution to be developed is mathematically exact only if the contact is perfectly lubricated or the materials considered are elastically similar, having the property
\begin{align}
 \frac{1 - 2\nu_{1}}{\mu_{1}} = \frac{1 - 2\nu_{2}}{\mu_{2}} \text{   ,} 
\end{align} 
where $\mu_{i}$ is the shear modulus and $\nu_{i}$ is the Possion's ratio of body $i$ for $i=1,2$. Although this limitation is not too restrictive for practical material combinations as they are commonly found in gas turbine engines, it is important to emphasise that the normal indentation does not induce tangential tractions. This means the problem is uncoupled and the solutions for the normal and shear tractions may be obtained independently.
 
The profile of the indenting punch is represented by the usual Hertz parabolic approximation to a circle with radius $R_{1}$ from $-a$ to $-b$, radius $R$ from $-b$ to $b$ and radius $R_{2}$ from $b$ to $c$. The relative surface normal displacement, $v(x)$, and its gradient, $v^{\prime }(x)$, can be expressed as piecewise functions

\begin{align}
v(x)=\left\{ 
\begin{array}{cc}
\Delta +\alpha x +\frac{x^{2}+2b k_{1} x+b^{2} k_{1} }{2R_{1}} & -a<x<-b \\ 
\Delta +\alpha x +\dfrac{x^{2}}{2R} & -b<x<b \\ 
\Delta + \alpha x +\frac{x^{2}-2b k_{2} x+b^{2} k_{2} }{2R_{2}} & b\leq x \leq c%
\end{array}%
\right. \Rightarrow v^{\prime }\left( x\right) =\left\{ 
\begin{array}{cc}
\alpha + \left( x+b k_{1}\right) /R_{1} & -a<x<-b \\ 
\alpha + x/R & -b\leq x \leq b \\ 
\alpha + \left( x-b k_{2}\right) /R_{2} & b<x<c%
\end{array}%
\right. ,
\end{align}
where $k_{i}=1-R_{i}/R$ for $i=1,2$. The absolute indentation is the constant $\Delta$. The inclination of the punch, $\alpha$, must be sufficiently small for small strain conventional elasticity to apply.

The normal traction distribution, $p(x)$, along the contact interface $[-a,c]$ can be found through inverting the following integral equation 
\begin{align}\label{integral_equation}
 \frac{1}{A} v^{\prime }(x) = \frac{1}{\uppi} \int_{-a}^{c} \frac{p(t) \, \mathrm{d}t }{(t - x)} \text{   .} 
\end{align}

Here, $A$ is the material `compliance' of the two touching bodies, which in a plane-strain formulation is
\begin{align}
 A = \frac{\kappa_1 +1}{4 \mu_{1}} + \frac{\kappa_2 +1}{4 \mu_{2}} \text{   ,} 
\end{align}
where $\kappa_i$ is Kolosov's constant, given by $\kappa_i = (3-4 \nu_i)$ for $i=1,2$.

\section{Normal indentation problem}	
\hspace{0.4cm}Equation \eqref{integral_equation} is a Cauchy singular integral equation of the first kind. In order to obtain the normal pressure distribution, $p(x)$, the integral equation  needs to be inverted using the theory of the Riemann-Hilbert boundary value problem for a non-closed contour. As the contact is incomplete in its nature, a bounded-bounded solution is expected and the following inversion formula follows
\begin{align}\label{inverted_IE}
p(x) = - \frac{w(x)}{\uppi A}  \int_{-a}^{c} \frac{ \, v^{\prime }(t) \, \mathrm{d}t }{\sqrt{(t+a)(c-t)}\,(t - x)}\text{   ,}
\end{align}
where $w(x)=\sqrt{(x+a)(c-x)} $  \cite{Hills_1993}. \\

Carrying out the integration, the pressure distribution is obtained as
\begin{align} \label{pressure_abbreviated}
p\left( x\right) =-\frac{w(x)}{\uppi A}\left( F_{0} +\frac{\alpha R_{1} + b k_{1}+x }{R_{1}\, w(x) }\, F_{1}(x) +\frac{\alpha R+ x}{R\, w(x)}\, F_{2}(x)  -\frac{-\alpha R_{2} + bk_{2}-x }{R_{2}\, w(x)}\, F_{3}(x)
\right)  \text{   ,}
\end{align}
where the expression $F_{0}$ and the functions $F_{j} (x)$, for $j=1,2,3$, as well as details of solving equation \eqref{inverted_IE} may be found in the appendix \ref{appendix:integral_evaluation}. \\

The solution is subject to the consistency condition
\begin{align}\label{consistency_cond}
\int_{-a}^{c} \frac{v^{\prime }(t) \, \mathrm{d}t }{\sqrt{(t+a)(c-t)}} = 0 \text{   .}
\end{align}

Evaluating the integral in equation \eqref{consistency_cond} (for details see appendix \ref{appendix:consistency}) gives an explicit expression for the inclination 

\begin{align}\label{consistency_condition}
\alpha =&\, \frac{1}{2 \pi }\left({}+{}\frac{k_1}{R_1} (a-2 b-c) \, \arccos\left[\frac{-a+2 b+c}{a+c}\right]+\frac{k_2}{R_2} (a+2 b-c) \, \arccos\left[\frac{a+2 b-c}{a+c}\right]{}+{} \right. \nonumber \\
& \quad \quad \left. \, \,  {}+{}\frac{\pi \,(a-c) }{R}+ \frac{k_1}{R_1}\, 2\, \sqrt{(a-b) (b+c)}-\frac{k_2}{R_2} \, 2\,  \sqrt{(a+b) (c-b)}\, \right)  \text{   .}
\end{align}

Finally, normal and rotational equilibrium are imposed as follows
\begin{align}
P  =&- \int\limits_{-a}^{c} p(x) \, \mathrm{d}x \text{} \label{vertical} \text{   ,} \\
P s  =& \int\limits_{-a}^{c} p(x)\, x \, \mathrm{d}x \text{   .} \label{rotational}
\end{align}

When evaluating the integrals stated above, the vertical equilibrium equation \eqref{vertical} gives rise to
\begin{align}\label{equilibrium_condition_1}
P=& \frac{1}{8 A} \left((a+c)^2 \left(-\frac{k_1}{R_1} \arccos\left[\frac{-a+2 b+c}{a+c}\right]-\frac{k_2}{R_2} \arccos\left[\frac{a+2 b-c}{a+c}\right]-\frac{\pi }{R}\right)- \right.  \nonumber \\ 
&\quad \quad \,\, \left. {}-{}\frac{k_1}{R_1}\, 2\, (a-2 b-c) \sqrt{(a-b) (b+c)}+\frac{k_2}{R_2}\, 2 \, (a+2 b-c) \sqrt{(a+b) (c-b)}  \right) \text{   ,}
\end{align}
and the rotational equilibrium equation \eqref{rotational} gives rise to
\begin{align}\label{equilibrium_condition_2}
Ps=& \frac{(a-c)}{48 A}\left(3 (a+c)^2 \left(\frac{k_1}{R_1} \arccos\left[\frac{-a+2 b+c}{a+c}\right]+\frac{k_2}{R_2} \arccos\left[\frac{a+2 b-c}{a+c}\right]+\frac{\pi }{R}\right) + \right. \nonumber \\
&\quad \quad \quad \quad {}+{} \frac{k_1}{R_1} \,4  \, \left(\frac{a^2-2 b^2+c^2}{a-c}+\frac{1}{2} (a-2 b-c)\right) \sqrt{(a-b) (b+c)} {}-{} \nonumber \\
&\quad \quad \quad \quad \left. -\frac{k_2}{R_2} \, 4 \, \left(\frac{a^2-2 b^2+c^2}{a-c}+\frac{1}{2} (a+2 b-c)\right) \sqrt{(a+b) (c-b)}\right)   \text{   .} 
\end{align}
The integration procedure for the equations stated above is given in  appendices \ref{appendix:vertical_eqilibirum} and \ref{appendix:rotational_eqilibirum}.

Equations \eqref{pressure_abbreviated}, \eqref{consistency_condition}, \eqref{equilibrium_condition_1}, and \eqref{equilibrium_condition_2} constitute a complete solution for the pressure distribution, inclination, and contact law. The equilibrium conditions suffice to determine the size of the contact interface $[-a,c]$.  For the reader's convenience, a complete solution including normalised plots of the pressure distribution is provided in the form of a Mathematica file, which may be found in the download section. 

\section{Results}\label{section_results}

\hspace{0.4cm}As a first check of the solution found the Hertzian case \cite{Hertz_1881} is recovered for the pressure distribution, $p(x)$, if the following limits are taken, i.e. $R_{2} \to R_{1}$, $b \to 0$, $c \to a$, and , $\alpha \to 0$ so that
 \begin{align}\label{pressure_distribution_Hertz}
 p(x)=&-\frac{\sqrt{a^2 - x^2}}{A R_{1}}  \text{   .}
 \end{align}
 
Also, the Hertzian contact law \cite{Hertz_1881},
 \begin{align}\label{contact_law_Hertz}
 \frac{P A}{R_{1}}=&-\,\frac{\uppi a^2}{2 R_{1}^2}  \text{   ,}
  \end{align}
is recovered if the limits $R_{2} \to R_{1}$, $b \to 0$, and $c \to a$ are taken for normal equilibrium given in equation \eqref{equilibrium_condition_1}.
  
A question demanding attention is under which circumstances results are in accordance with half-plane theory. In cases where the ratios $(a-b)/R_1$ and $(c-b)/R_2$ are small, i.e. when the contact zone extends only a small way into the outer curved parts of the indenter, the remaining amount of material outside of the contact would certainly suffice for the punch to be thought of as a half-plane. As the contact interface extends further into the outer regions, the approximation of the curved parts as a parabola must be called into question, and the half-plane assumption will also become deficient. On the other hand, if the central segment, with $R$ and $2b$, becomes very large compared to the radii $R_1$ and $R_2$, i.e. the edges of the punch appear to be very sharp, half-plane theory becomes increasingly strained regardless of the extent of the outer parts in contact.

\begin{figure}[H]
	\centering
	\includegraphics[scale=0.32, trim=20 1930 0 80, clip]{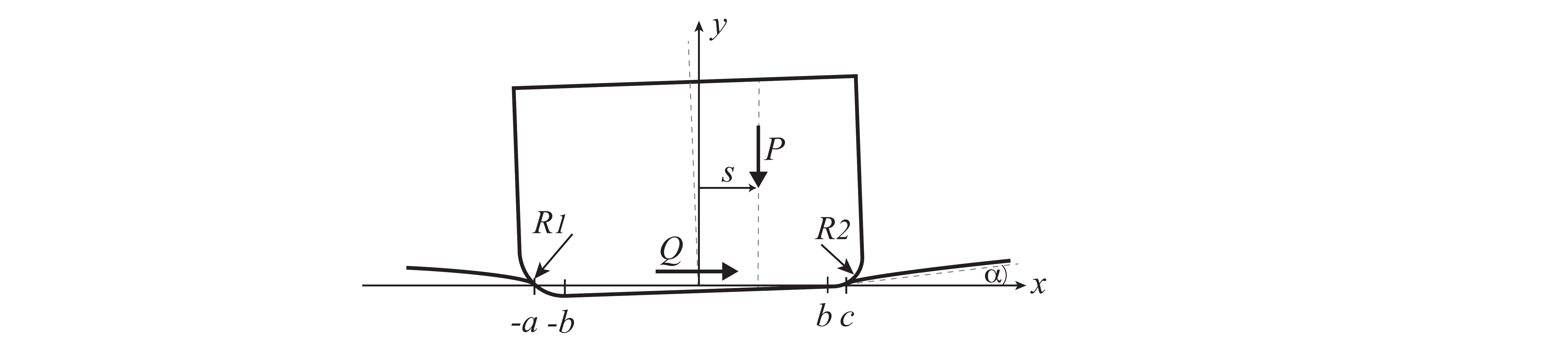}
	\caption{Illustration of the tilted flat but rounded punch with unequal radii}
	\label{fig:illustration_case_CC}	
\end{figure}

It appears cumbersome, if not impossible, to present all possible parameter combinations within this work in a comprehensive way. This is why we decided to take the limit $R \to \infty$ and explore case `A' as shown in figure \ref{fig:illustration_case_CC} in more detail. As described in the introduction of this work the tilted flat but rounded punch is a representation of a dovetail geometry, which is of particular interest to the aeronautic industries. 


There is a multitude of possibilities to illustrate the pressure distribution. The most sensible way seems to normalise the contact pressure with respect to $P/(2b)$ and choose a dimensionless load, $4PA/b = 1.0$, such that changes due to an increase in moment or a decrease in the ratio $R_2/R_1$ become most obvious. In order to examine the effect of a varying lever arm, $s$, we restrict the problem, first, to a symmetrical indenter, hence ${R_2}$/${R_1} = 1$. The presented results were obtained for a solid material with a Young's modulus of $E=210,000\,\nicefrac{\text{N}}{\text{mm}^2}$ and Possion's ratio of $\nu=0.3$.

\begin{figure}[H]
	\centering
	\includegraphics[scale=0.58, trim=0 370 0 0, clip]{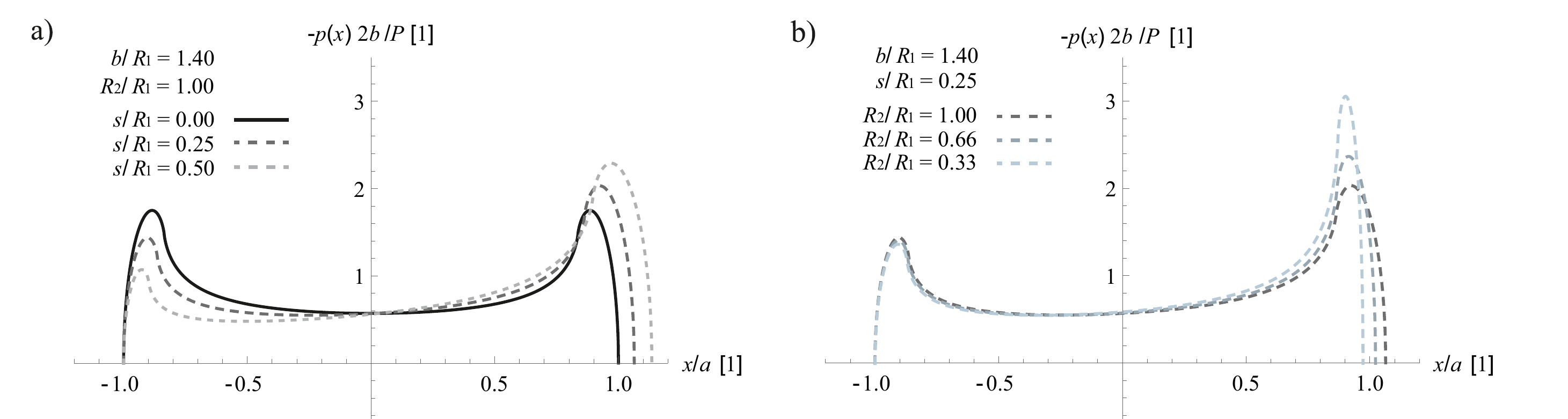}
	\caption{Comparison of normal pressure distributions for a) different lever arms, $s$, and b) different ratios of $R_{2}/R_{1}$ when, additionally, a constant clockwise moment is applied}
	\label{fig:comparison_+moment}	
\end{figure}
Figure \ref{fig:comparison_+moment} a) shows that the peak pressure decreases at the left edge and increases at the right edge as the clockwise moment grows. Also, the asymmetry of the contact interface increases as the moment grows. In Figure  \ref{fig:comparison_+moment} b) the moment is kept constant with a lever arm of $s/R_1=0.25$. As the ratio of $R_2/R_1$ is decreased it becomes apparent how significant the differences in the peak pressures at each end of the contact may be if the radii are unequal and the normal loading leans towards the sharper edge. However, the contact pressure at the left edge, where the radius does not change, appears to stay almost the same. This insight may be helpful when employing asymptotic solutions to asymmetrical contacts.
 
For industry applications it must be emphasised that neither the application of a moment nor unequal radii result in a significant asymmetry in the contact patch, i.e. the ratio of contact coordinates, ${a}/{c}$, is almost constant for applicable parameter combinations. On the other hand, the difference in peak pressure at the respective edges is significant.  

\begin{figure}[H]
	\centering
	\includegraphics[scale=0.27, trim=0 200 0 10, clip]{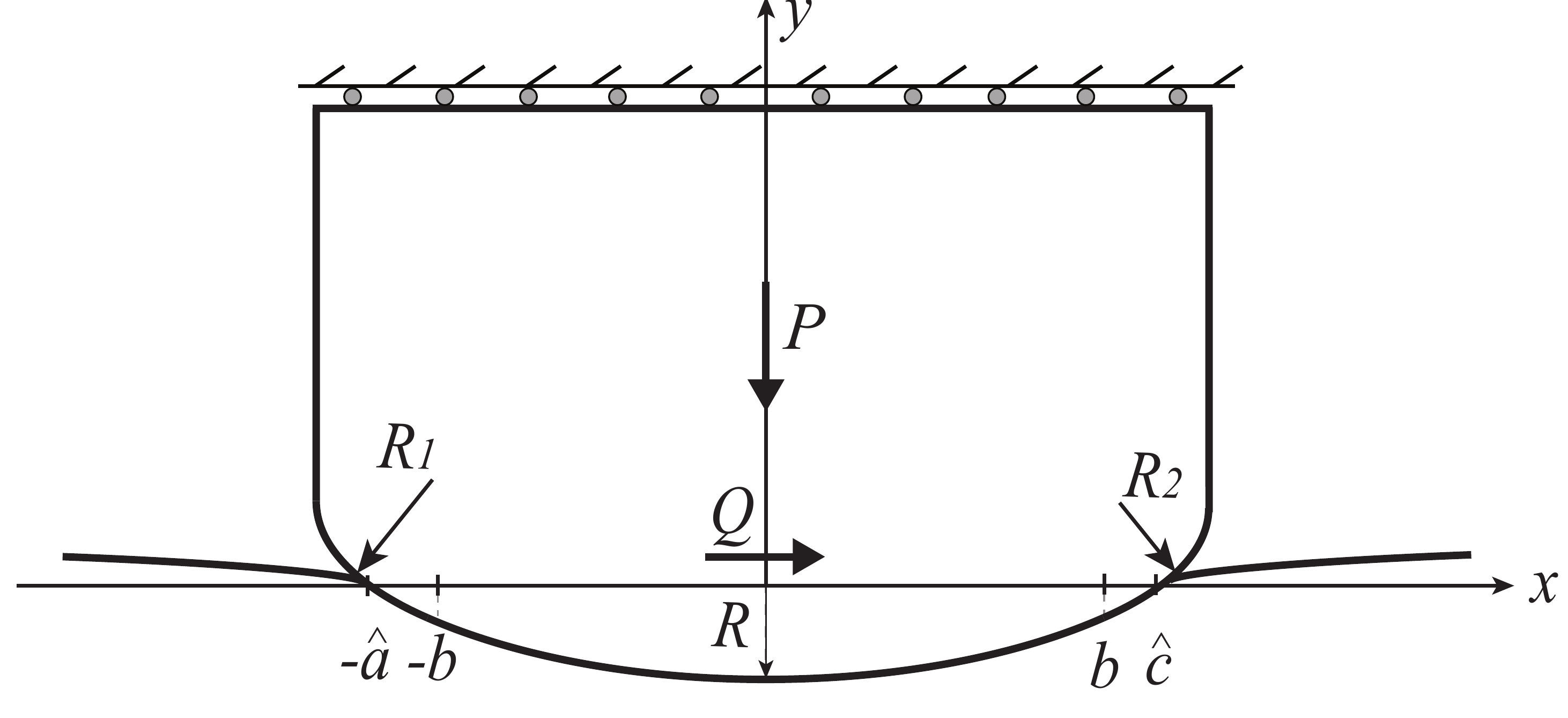}
	\caption{Illustration of the three-segment punch with unequal radii without tilt}
	\label{fig:illustration_case_CCC}	
\end{figure}

Another interesting remark regards a change in boundary conditions of the problem statement as depicted in figure \ref{fig:illustration_case_CCC}. Suppose one is interested in a contact interface, which develops when any inclination, $\alpha$, of the punch is suppressed through a support structure. This means the asymmetrical punch is pressed vertically into the other half-plane resulting in an asymmtrical contact interface and pressure distribution different from what was examined in the example above. However, in such a case the algebraic expressions for the pressure distribution and normal equilibrium given in equations \eqref{pressure_abbreviated} and \eqref{equilibrium_condition_1} do not change. The new contact interface $[-\hat{a},\hat{c}]$ can be determined by imposing equation \eqref{consistency_condition} to equal zero, $\alpha=0$, and evaluating the consistency condition and normal equilibrium numerically for $\hat{a}$ and $\hat{c}$. Suppose the support structure is removed and we would like to apply an external moment, $P s^{\ast}$, where the lever arm, $s^{\ast}$, is chosen such that it suppresses any tilt of the punch due to the geometrical asymmetry. We can find this lever arm by evaluating the rotational equilibrium given in equation \eqref{equilibrium_condition_2} for the contact interface $[-\hat{a},\hat{c}]$ and solving for $s^{\ast}$.

Returning to the original boundary conditions, the solution framework found requires $0 \leq (a-b)$ and $0 \leq (c-b)$. A further question to answer is which maximum moment can be applied to cause $c=b$. That is, one of the outer rounded parts has lifted out of contact and the contact interface on one side ends at the transition point of the central segment and the outer rounded part, whereas $0 \leq (a-b)$ is still satisfied. In order to obtain a compact algebraic expression the limit $c \to b$ must be taken. The normal equilibrium gives
\begin{align}\label{Pctob}
P_{\text{lift-off}} =&\frac{1}{8 A}\left((a+b)^2 \left(-\frac{k_1 }{R_1} \arccos \left[\frac{3 b-a}{a+b}\right]-\frac{\pi }{R}\right)-\frac{ k_1}{R_1} \, 2 \, (a-3 b) \sqrt{2 b (a-b)}\right)  \text{  .}
\end{align}

When taking the limit $c \to b$ for the rotational equilibrium, substituting the load, $P_{\text{lift-off}}$, with the expression obtained in equation \eqref{Pctob}, and solving for the lever arm, $s_{\text{lift-off}}$, it follows

\begin{align}
s_{\text{lift-off}} = &-\frac{ (a-b)\,R \left(\frac{2 \sqrt{2} b k_1}{\sqrt{b (a-b)}} \left(3 a^2-4 a b+3 b^2\right)+3 (a+b)^2 \left(k_1 \arccos\left[-\frac{a-3 b}{a+c}\right]+\frac{\pi  R_1}{R} \right)\right)}{6 \left((a+b)^2 \left(k_1 R \arccos\left[3-\frac{4 a}{a+b}\right]+\pi R_1 \right)+2 \sqrt{2} k_1 R (a-3 b) \sqrt{b (a-b)}\right)} \text{   .}
\end{align}	 
 
\section{Application of a monotonically increasing shear force}

\hspace{0.4cm}  Suppose, now, that there is friction between the punch and the half-plane with a finite coefficient of friction, $f$, and the contacting bodies are elastically similar. The subsequent application of a monotonically increasing shear force, $Q$, produces either a state of sliding, given Coulomb's friction law applies, if $\vert Q\vert = f P$, or a partial slip pattern develops, when the inequality $\vert Q\vert < f P$ holds. The latter provokes the immediate question how large the stick and slip zones in the contact interface may be.

\begin{figure}[H]
	\centering
	\includegraphics[scale=0.35, trim=100 1870 0 70, clip]{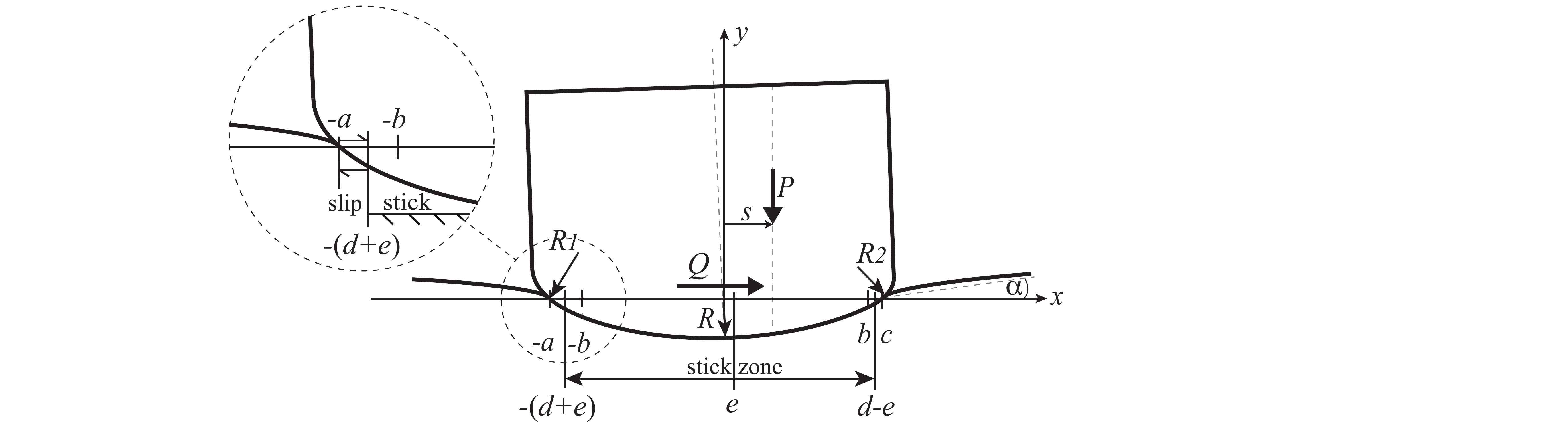}
	\caption{Illustration of the tilted punch comprised of three different rounded segments in partial slip}
	\label{fig:illustration_tilted_punch_partial_slip}	
\end{figure}

\hspace{0.4cm}An answer to this question can be found using the Ciavarella-J\"ager theorem \cite{Ciavarella_1998}, where the shear traction is viewed as the sum of the sliding shear tractions, $-f p(x)$, and a corrective shear traction distribution, $q^{\ast} (x)$, in the stick zone
\begin{align}
q(x) =\left\{ 
\begin{array}{cll}
&-f p(x)  & 
\text{ , }-a\leq x\leq -(d+e)\\
&-f p(x) +q^{\ast}(x)  & 
\text{ , }\left\vert x+e\right\vert <d \\ 
&-f p(x)  & 
\text{ , } (d-e) \leq x\leq c 
\end{array}%
\right. \text{   ,}
\end{align}
where the extent of the stick zone is denoted by $[-(d+e),\, (d-e)]$. Here, $e$ is the eccentricity of stick zone due to the asymmetric contact interface. Figure \ref{fig:illustration_tilted_punch_partial_slip} illustrates that a stick zone of size $[-(d+e),(d-e)]$ will remain in the contact interface $[-a,c]$. The points outside of the stick zone will slip and cause differential surface displacements of same sign at each end of the contact interface.

For reasons which will become clear, we write $q^{\ast} (x) = f p^{\ast} (x)$, where $p^{\ast}(x)$ is yet unknown. Suppose we are now interested in the integral relating to tangential tractions, which gives the relative difference in tangential surface strains, $\frac{\mathrm{d}g}{\mathrm{d}x} = \varepsilon_{xx1}(x) - \varepsilon_{xx2}(x)$, as
\begin{align}\label{integral_equation_tangential}
\frac{1}{A} \frac{\mathrm{d}g}{\mathrm{d}x} = \frac{f}{\uppi} \int_{-a}^{c} \frac{p(t) \, \mathrm{d}t }{(t - x)} + \frac{f}{\uppi} \int_{-d+e}^{d-e} \frac{p^{\ast}(t) \, \mathrm{d}t }{(t - x)} \text{   .} 
\end{align}

We further impose the requirement that the strain difference vanishes in the stick zone, i.e.
\begin{align}\label{stick_zone_strain}
\frac{\mathrm{d}g}{\mathrm{d}x} = 0 \quad ,\, -(d+e) < x < (d-e) \text{   .} 
\end{align}

Referencing back to equation \eqref{integral_equation}, we now see that the integral relating to the normal traction becomes
\begin{align}\label{integral_equation_2}
\frac{1}{A} v^{\prime }(x) = \frac{1}{\uppi} \int_{-(d+e)}^{d-e} \frac{p^{\ast}(t) \, \mathrm{d}t }{(t - x)} \text{   .} 
\end{align}

This reveals that the corrective normal traction, $p^{\ast}(x)$, and hence the corrective shear traction $q^{\ast}(x) = f p^{\ast}(x)$, might be obtained by simply substituting $a=(d+e)$ and $c=(d-e)$ in the pressure distribution given in equation \eqref{pressure_abbreviated}. However, we want to quantify the extent of the stick zone by determining the two unknowns $d$ and $e$. For that we will need two side conditions. One can be found by imposing horizontal equilibrium. Here, the following distinction between the shear force in full sliding, $Q_{fP}$, and the corrective term, $Q^{\ast}$, is made
\begin{align}\label{horizontal_equilibrium}
Q= Q_{fP} + Q^{\ast} = -f P + \int_{-(d+e)}^{d-e} q^{\ast}(x) \,\mathrm{d}x \, \text{.}
\end{align} 
The magnitude of the corrective shear force is determined by
\begin{align}
Q^{\ast} = \vert \frac{Q}{f P} - 1 \vert \, f P \text{.}
\end{align}

An algebraic expression for the last term in equation \eqref{horizontal_equilibrium} can be found immediately by substituting $a=(d+e)$ and $c=(d-e)$ in the equilibrium condition given in equation  \eqref{equilibrium_condition_1}. The other necessary side condition is found by carrying out the same substitution for the consistency condition given in equation \eqref{consistency_condition}. Now, horizontal equilibrium and the consistency condition for the tangential tractions suffice to numerically determine $d$ and $e$.\\

\begin{figure}[H]
	\centering
	\includegraphics[scale=0.58, trim=0 370 0 0, clip]{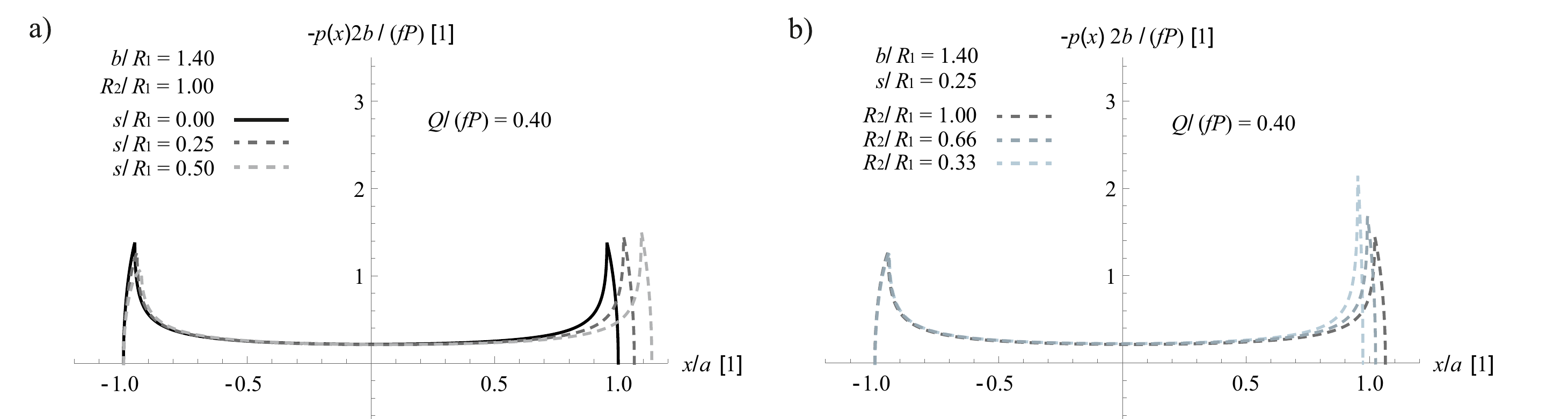}
	\caption{Comparison of shear pressure distributions in partial slip ($Q/(fP) = 0.4$) for a) different lever arms, $s$, and b) different ratios of $R_{2}/R_{1}$ when, additionally, a constant clockwise moment is applied}
	\label{fig:comparison_shear_pressure}	
\end{figure}

Figure \ref{fig:comparison_shear_pressure} depicts a comparison of shear pressure distributions in partial slip ($Q/(fP) = 0.4$) for a) different lever arms, $s$, and b) different ratios of $R_{2}/R_{1}$ when the moment is kept constant at $s/R_1=0.25$.  The given example is resting on the same inputs as the results presented for the normal load problem considered in section \ref{section_results}.

Obviously, the shear tractions in full sliding are related to the normal tractions by $q(x) = f p(x)$. The partial slip solution is restricted in a sense that the stick zone needs to be bounded by the outer rounded parts of the three-part indenter and, more importantly, cannot end in the central segment as this violates the substitution of $a=(d+e)$ and $c=(d-e)$.

\section{Conclusions}

\hspace{0.4cm} This paper's content is threefold. Firstly, a complete description is provided of the normal indentation for a generic three-segment piecewise-quadratic half-plane contact subject to a normal load and moment. Secondly, the application of a monotonically increasing shear force and the developing partial slip pattern is studied using the Ciavrella-J\"ager theorem. Lastly, based on the comprehensive solution found, three specific indenter problems are derived. Thorough focus is given to the tilted flat and rounded punch with unequal radii since it is capable of representing a dovetail flank contact in a realistic manner. It is shown that the contact pressure at the sharper edge can be significantly higher than in the symmetrical case, whereas the effect of the geometrical asymmetry at the sharper edge does not affect the pressure distribution on the other end of the contact significantly. For the reader's convenience a set of Mathematica files is provided in order to comprehensively illustrate the found solutions and to facilitate further research activities. 

The insights gained through this work shall set the basis for further investigations of complex loading cycles, including varying normal and tangential forces as well asymptotic methods in order to match simpler laboratory experiments. Furthermore, the study of the evolution of frictional energy dissipation in asymmetrical contacts is necessary in order to determine damping properties of the contact interface and assessing the localisation of surface damage which may give rise to fretting damage and possible crack nucleation.

\section*{Acknowledgments}
This project has received funding from the European Union's Horizon 2020 research and innovation programme under the Marie Sklodowska-Curie agreement No 721865.

\bibliographystyle{unsrt}

\newpage

\begin{appendix}

\numberwithin{equation}{section}
\section{Appendix}
\label{chp:Appendix}

\subsection{Integral evaluation}\label{appendix:integral_evaluation}

As the punch's profile, $v(x)$, (and also its derivative, $v^{\prime}(x)$) is a piecewise function, also the normal surface stress can be expressed as a piecewise function. Hence, the normal pressure can be written as
\begin{align}
p\left( x\right) =\dfrac{w(x)}{\pi A} \left\{ 
\begin{tabular}{l}
$-\dfrac{1}{R_{1}}\left[ \int\limits_{-a}^{-b}\dfrac{%
\left(\alpha R_{1} + t+bk_{1}\right)\,\mathrm{d}t }{w(t)\left( t-x\right) }\right]^{\text{\tiny CPV}} -\dfrac{%
1}{R}\int\limits_{-b}^{b}\dfrac{(\alpha R + t)\, \mathrm{d}t}{w(t)\left(
t-x\right)}-\dfrac{1}{R_{2}}\int\limits_{b}^{c}\dfrac{%
\left(\alpha R_{2} + t-bk_{2}\right)\, \mathrm{d}t }{w(t)\left( t-x\right) },$ $-a<x<-b$ \\ 
$-\dfrac{1}{R_{1}}\int\limits_{-a}^{-b}\dfrac{\left(
\alpha R_{1} + t+bk_{1}\right)\, \mathrm{d}t }{w(t)\left( t-x\right) }-\dfrac{1}{R}\left[ \int\limits_{-b}^{b}\dfrac{(\alpha R +t)\, \mathrm{d}t}{w(t)%
\left(t-x\right) }\right]^{\text{\tiny CPV}} -\dfrac{1}{R_{2}}%
\int\limits_{b}^{c}\dfrac{\left(\alpha R_{2} +t-bk_{2}\right)\, \mathrm{d}t }{w(t)\left( t-x\right) }%
,$ $-b<x<b$ \\ 
$-\dfrac{1}{R_{1}}\int\limits_{-a}^{-b}\dfrac{\left(
\alpha R_{1} +t+bk_{1}\right)\, \mathrm{d}t }{w(t)\left( t-x\right) }-\dfrac{1}{R}\int\limits_{-b}^{b}\dfrac{(\alpha R +t)\, \mathrm{d}t}{w(t)\left(
t-x\right) }-\dfrac{1}{R_{2}} \left[ \int\limits_{b}^{c}%
\dfrac{\left(\alpha R_{2} +t-bk_{2}\right)\, \mathrm{d}t }{w(t)\left( t-x\right) }\right]^{\text{\tiny CPV}} ,$
$b<x<c$%
\end{tabular}%
\right. \text{   ,}
\end{align}
where $w(x)=\sqrt{(x+a)(c-x)} $ and $w(t)=\sqrt{(t+a)(c-t)}$, such that the integrals can be evaluated. The integrals marked with CPV are singular and must be evaluated as Cauchy principal values (CPV).\\

First, a partial decomposition on the integrands is performed so that
\begin{align}
\dfrac{\left(\alpha R_{1} + t+bk_{1}\right) }{\left( t-x\right) } &=1+\frac{\alpha R_{1} + bk_{1}+x}{t-x} \text{   ,}
\\
\dfrac{\left(\alpha R + t\right)}{\left( t-x\right) } &=1+\frac{\alpha R + x}{t-x}  \text{   , and}
\\
\dfrac{\left(\alpha R_{2} + t-bk_{2}\right) }{\left( t-x\right) } &=1-\frac{-\alpha R_{2} + bk_{2}-x}{t-x}  \text{   .}
\end{align}

The integrals that are regular and independent of $x$ can be evaluated as follows
\begin{align}
\int_{-a}^{-b}\frac{1 \, \mathrm{d}t}{w\left( t\right) } & =\arccos\left[\frac{2 b-(a-c)}{a+c}\right]  \text{   ,} \\
\int_{-b}^{b}\frac{1 \, \mathrm{d}t}{w\left( t\right) } & =\arccos\left[\frac{2 b-(a-c)}{a+c}\right] + \arccos\left[\frac{(a-c)+2 b}{a+c}\right] - \pi   \text{   , and} \\
\int_{b}^{c}\frac{1 \, \mathrm{d}t}{w\left( t\right) } & =\arccos\left[\frac{(a-c)+2 b}{a+c}\right] \text{   .}
\end{align}

The contribution of these integrals to the normal pressure is
\begin{align}\label{f0}
F_{0} = \frac{k_1}{R_1} \arccos\left[\frac{2 b-(a-c)}{a+c}\right]+\frac{k_2}{R_2} \arccos\left[\frac{(a-c)+2 b}{a+c}\right]+\frac{\pi }{R}	\text{   .}
\end{align}

\newpage
The results for the remaining integrals including those affected by the CPV are the following

\begin{align}
\text{CPV} \int_{-a}^{-b}\frac{1}{w\left( t\right) \left( t-x\right) } \mathrm{d}t &= +\frac{%
\ln \left( -\frac{\left( b+x\right) \left( c+a\right) }{2\sqrt{\left(
c-x\right) \left( a+x\right) \left( a-b\right) \left( c+b\right) }+2\left(
ac+bx\right) +\left( a-c\right) \left( b-x\right) }\right) }{\sqrt{\left(
c-x\right) \left( a+x\right) }},\text{ }x\in \left[ -a,-b\right] \text{   ,}  \\
\int_{-a}^{-b}\frac{1}{w\left( t\right) \left( t-x\right) }\mathrm{d}t &=+\frac{\ln
\left( \frac{\left( b+x\right) \left( c+a\right) }{2\sqrt{\left( c-x\right)
\left( a+x\right) \left( a-b\right) \left( c+b\right) }+2\left( ac+bx\right)
+\left( a-c\right) \left( b-x\right) }\right) }{\sqrt{\left( c-x\right)
\left( a+x\right) }},\text{ }x\notin \left[ -a,-b\right]  \text{   ,} \\
\text{CPV}\int_{-b}^{b}\frac{1}{w\left( t\right) \left( t-x\right) }\mathrm{d}t &=+\frac{\ln
\left( \frac{\left( 2\sqrt{\left( c-x\right) \left( a+x\right) \left(
a-b\right) \left( c+b\right) }+2\left( ac+bx\right) +\left( a-c\right)
\left( b-x\right) \right) \left( b-x\right) }{\left( 2\sqrt{\left(
c-x\right) \left( a+x\right) \left( a+b\right) \left( c-b\right) }+2\left(
ac-bx\right) -\left( a-c\right) \left( b+x\right) \right) \left( b+x\right) }%
\right) }{\sqrt{\left( c-x\right) \left( a+x\right) }},\text{ }x\in \left[
-b,b\right] \text{   ,}  \\
\int_{-b}^{b}\frac{1}{w\left( t\right) \left( t-x\right) }\mathrm{d}t &=+\frac{\ln
\left( -\frac{\left( 2\sqrt{\left( c-x\right) \left( a+x\right) \left(
a-b\right) \left( c+b\right) }+2\left( ac+bx\right) +\left( a-c\right)
\left( b-x\right) \right) \left( b-x\right) }{\left( 2\sqrt{\left(
c-x\right) \left( a+x\right) \left( a+b\right) \left( c-b\right) }+2\left(
ac-bx\right) -\left( a-c\right) \left( b+x\right) \right) \left( b+x\right) }%
\right) }{\sqrt{\left( c-x\right) \left( a+x\right) }},\text{ }x\notin \left[
-b,b\right]  \text{   ,} \\
\text{CPV}\int_{b}^{c}\frac{1}{w\left( t\right) \left( t-x\right) }\mathrm{d}t &=-\frac{\ln
\left( -\frac{\left( b-x\right) \left( c+a\right) }{2\sqrt{\left( c-x\right)
\left( a+x\right) \left( a+b\right) \left( c-b\right) }+2\left( ac-bx\right)
-\left( a-c\right) \left( b+x\right) }\right) }{\sqrt{\left( c-x\right)
\left( a+x\right) }},\text{ }x\in \left[ b,c\right] \text{   , and}  \\
\int_{b}^{c}\frac{1}{w\left( t\right) \left( t-x\right) }\mathrm{d}t &=-\frac{\ln
\left( \frac{\left( b-x\right) \left( c+a\right) }{2\sqrt{\left( c-x\right)
\left( a+x\right) \left( a+b\right) \left( c-b\right) }+2\left( ac-bx\right)
-\left( a-c\right) \left( b+x\right) }\right) }{\sqrt{\left( c-x\right)
\left( a+x\right) }},\text{ }x\notin \left[ b,c\right] \text{   .}
\end{align}

Therefore, regardless whether they are CPV or not, they can be written as
\begin{align}
\int_{-a}^{-b}\frac{1}{w\left( t\right) \left( t-x\right) }\mathrm{d}t =&+\frac{\ln
\left( \left\vert \frac{\left( b+x\right) \left( c+a\right) }{2\sqrt{\left(
c-x\right) \left( a+x\right) \left( a-b\right) \left( c+b\right) }+2\left(
ac+bx\right) +\left( a-c\right) \left( b-x\right) }\right\vert \right) }{%
\sqrt{\left( c-x\right) \left( a+x\right) }} = \frac{F_{1}(x)}{w(x)}  \text{   ,} \\
\int_{-b}^{b}\frac{1}{w\left( t\right) \left( t-x\right) }\mathrm{d}t =&+\frac{\ln
\left( \left\vert \frac{\left( 2\sqrt{\left( c-x\right) \left( a+x\right)
\left( a-b\right) \left( c+b\right) }+2\left( ac+bx\right) +\left(
a-c\right) \left( b-x\right) \right) \left( b-x\right) }{\left( 2\sqrt{%
\left( c-x\right) \left( a+x\right) \left( a+b\right) \left( c-b\right) }%
+2\left( ac-bx\right) -\left( a-c\right) \left( b+x\right) \right) \left(
b+x\right) }\right\vert \right) }{\sqrt{\left( c-x\right) \left( a+x\right) }%
} = \frac{F_{2}(x)}{w(x)} \text{   , and} \\
  \int_{b}^{c}\frac{1}{w\left( t\right) \left( t-x\right) }\mathrm{d}t =&-\frac{\ln
  \left( \left\vert \frac{\left( b-x\right) \left( c+a\right) }{2\sqrt{\left(
  c-x\right) \left( a+x\right) \left( a+b\right) \left( c-b\right) }+2\left(
  ac-bx\right) -\left( a-c\right) \left( b+x\right) }\right\vert \right) }{%
  \sqrt{\left( c-x\right) \left( a+x\right) }} = \frac{F_{3}(x)}{w(x)}  \text{   .}
\end{align}

\subsection{The consistency condition}\label{appendix:consistency}

The consistency condition can be evaluated as follows
\begin{align}\label{consistency_integrals}
0 &= \int_{-a}^{c} \frac{v^{\prime }(t) \, \mathrm{d}t }{\sqrt{(t+a)(c-t)}} \nonumber \\
&=\frac{1}{R_{1}} \int_{-a}^{-b} \frac{ \left(\alpha R_{1} + (x+ b k_{1})\right) \, \mathrm{d}t }{\sqrt{(t+a)(c-t)}} + \frac{1}{R} \int_{-b}^{b} \frac{ \left(\alpha R + x \right) \, \mathrm{d}t }{\sqrt{(t+a)(c-t)}} + \frac{1}{R_{2}} \int_{b}^{c} \frac{ \left(\alpha R_{2} + (x - b k_{2})\right) \, \mathrm{d}t }{\sqrt{(t+a)(c-t)}} \text{   .}
\end{align}

Evaluating the integrals in equation \eqref{consistency_integrals} and solving the resulting equation for $\alpha$ gives the algebraic expression presented in equation \eqref{consistency_condition}.

\subsection{The normal equilibrium condition}\label{appendix:vertical_eqilibirum}

Normal equilibrium must be satisfied so that
\begin{align}
P  =&- \int\limits_{-a}^{c} p(x) \, \mathrm{d}x \nonumber \\
=&- \int\limits_{-a}^{c} \left[ - \frac{1}{\pi A}  w(x) \int\limits_{-a}^{c} \frac{v^{\prime}(t) \mathrm{d}t}{w(t) (t-x)}\right ] \mathrm{d}x \nonumber \\
=&+ \frac{1}{\pi A} \int\limits_{-a}^{c} \left[  \frac{v^{\prime}(t)}{w(t)} \int\limits_{-a}^{c} \frac{w(x) \mathrm{d}x}{(t-x)}\right ] \mathrm{d}t \nonumber \\
=&+  \frac{1}{\pi A} \int\limits_{-a}^{c} \left[  \frac{v^{\prime}(t)}{w(t)} \left(\frac{1}{2} \pi (a-c+2 t)\right) \right ] \mathrm{d}t \nonumber \\
=&+  \frac{1}{2 A} \int\limits_{-a}^{c} \left[  \frac{(a-c+2 t)}{w(t)} v^{\prime}(t) \right ] \mathrm{d}t   \text{   .}
\end{align}

Integration by parts (where $f^{\prime}(t) = \frac{(a-c+2t)}{w(t)}$ and $g(t) = v^{\prime}(t)$) gives
\begin{align}
P  =&+  \frac{1}{2 A} \left[ \left[ f(t) g(t) \mathop{\big|}\limits_{-a}^{c} \right ]  - \int\limits_{-a}^{c} f(t) g^{\prime} (t) \mathrm{d}t \right ] \nonumber \\
=&+ \frac{1}{2 A} \left[+ \left( \left[ f(t) g(t) \mathop{\big|}\limits_{-a}^{-b} \right ]  - \int\limits_{-a}^{-b} f(t) g^{\prime} (t) \mathrm{d}t \right ) + \nonumber \right. \\
& \quad \quad \quad \, \left. \,+  \left( \left[ f(t) g(t) \mathop{\big|}\limits_{-b}^{b} \right ]  - \int\limits_{-b}^{b} f(t) g^{\prime} (t) \mathrm{d}t \right ) + \left( \left[ f(t) g(t) \mathop{\big|}\limits_{b}^{c} \right ]  - \int\limits_{b}^{c} f(t) g^{\prime} (t) \mathrm{d}t \right ) \right ] \text{   .}
\end{align}

The indefinite integral of  $f^{\prime}(t)$ is
\begin{align}
f(t) = \int f^{\prime}(t) \mathrm{d}t  = \int \frac{(a-c+2t)}{w(t)} \mathrm{d}t = -2 \sqrt{(a+t)(c-t)} + C_1  \text{   ,}
\end{align}
where $C_1$ is a constant.\\

The derivative of $g(t) =v^{\prime}(t)$ gives the piecewise function
\begin{align}
g^{\prime}(t) = v^{\prime \prime }\left( t\right) =\left\{  
\begin{array}{ll}
1/R_{1} & ,-a<t<-b \\ 
1/R & ,-b<t<b \\ 
1/R_{2} & ,\,b<t<c
\end{array}%
\right.  \text{   .}
\end{align}

Now normal equilibrium is 
\begin{align}
P=& \frac{1}{8 A} \left((a+c)^2 \left(-\frac{k_1}{R_1} \arccos\left[\frac{-a+2 b+c}{a+c}\right]-\frac{k_2}{R_2} \arccos\left[\frac{a+2 b-c}{a+c}\right]-\frac{\pi }{R}\right)- \right.  \nonumber \\ 
&\quad \quad \,\, \left. {}-{}\frac{k_1}{R_1}\, 2\, (a-2 b-c) \sqrt{(a-b) (b+c)}+\frac{k_2}{R_2}\, 2 \, (a+2 b-c) \sqrt{(a+b) (c-b)}  \right) \text{   .}
\end{align}

\subsection{The rotational equilibrium condition}\label{appendix:rotational_eqilibirum}

For rotational equilibrium the same procedure as for normal equilibrium applies
\begin{align}
P s  =& \int\limits_{-a}^{c} p(x)\, x \, \mathrm{d}x \nonumber \\
=& \int\limits_{-a}^{c} \left[ - \frac{1}{\pi A}  w(x) \, x\, \int\limits_{-a}^{c} \frac{v^{\prime}(t) \mathrm{d}t}{w(t) (t-x)}\right ] \mathrm{d}x \nonumber \\
=& - \frac{1}{\pi A} \int\limits_{-a}^{c} \left[  \frac{v^{\prime}(t)}{w(t)} \int\limits_{-a}^{c} \frac{w(x) \, x \,\mathrm{d}x}{(t-x)}\right ] \mathrm{d}t \nonumber \\
=& - \frac{1}{\pi A} \int\limits_{-a}^{c} \left[  \frac{v^{\prime}(t)}{w(t)} \left(\frac{1}{8} \pi  \left(4 t (a-c)-(a+c)^2+8 t^2\right)\right) \right ] \mathrm{d}t \nonumber \\
=& - \frac{1}{8 A} \int\limits_{-a}^{c} \left[  \frac{\left(4 t (a-c)-(a+c)^2+8 t^2\right)}{w(t)} v^{\prime}(t) \right ] \mathrm{d}t  \text{   .}
\end{align}

Integration by parts (where $f^{\prime}(t) = \frac{\left(4 t (a-c)-(a+c)^2+8 t^2\right)}{w(t)}$ and $g(t) = v^{\prime}(t)$) gives
\begin{align}
P s  =& - \frac{1}{8 A} \left[ \left[ f(t) g(t) \mathop{\big|}\limits_{-a}^{c} \right ]  - \int\limits_{-a}^{c} f(t) g^{\prime} (t) \mathrm{d}t \right ] \nonumber \\
=&  - \frac{1}{8 A} \left[ + \left( \left[ f(t) g(t) \mathop{\big|}\limits_{-a}^{-b} \right ]  - \int\limits_{-a}^{-b} f(t) g^{\prime} (t) \mathrm{d}t \right ) + \left( \left[ f(t) g(t) \mathop{\big|}\limits_{-b}^{b} \right ]  - \int\limits_{-b}^{b} f(t) g^{\prime} (t) \mathrm{d}t \right ) + \nonumber \right. \\
& \quad \quad \quad \, \left. \,+ \left( \left[ f(t) g(t) \mathop{\big|}\limits_{b}^{c} \right ]  - \int\limits_{b}^{c} f(t) g^{\prime} (t) \mathrm{d}t \right ) \right ]  \text{   .}
\end{align}

The indefinite integral of  $f^{\prime}(t)$ is
\begin{align}
f(t) = \int f^{\prime}(t) \mathrm{d}t  = \int \frac{\left(4 t (a-c)-(a+c)^2+8 t^2\right)}{w(t)} \mathrm{d}t = -2 \sqrt{(a+t) (c-t)} (-a+c+2 t) + C_2  \text{   ,}
\end{align}
where $C_2$ is a constant. \\

The derivative of $g(t) =v^{\prime}(t)$ gives the piecewise function
\begin{align}
g^{\prime}(t) = v^{\prime \prime }\left( t\right) =\left\{  
\begin{array}{ll}
1/R_{1} & ,-a<t<-b \\ 
1/R & ,-b<t<b \\ 
1/R_{2} & ,\,b<t<c
\end{array}%
\right.   \text{   .}
\end{align}

The rotational equilibrium condition gives
\begin{align}\label{rotationalequilibrium_condition}
Ps=& \frac{(a-c)}{48 A}\left(3 (a+c)^2 \left(\frac{k_1}{R_1} \arccos\left[\frac{-a+2 b+c}{a+c}\right]+\frac{k_2}{R_2} \arccos\left[\frac{a+2 b-c}{a+c}\right]+\frac{\pi }{R}\right) + \right. \nonumber \\
&\quad \quad \quad \quad {}+{} \frac{k_1}{R_1} \,4  \, \left(\frac{a^2-2 b^2+c^2}{a-c}+\frac{1}{2} (a-2 b-c)\right) \sqrt{(a-b) (b+c)} {}-{} \nonumber \\
&\quad \quad \quad \quad \left. -\frac{k_2}{R_2} \, 4 \, \left(\frac{a^2-2 b^2+c^2}{a-c}+\frac{1}{2} (a+2 b-c)\right) \sqrt{(a+b) (c-b)}\right)   \text{   .} 
\end{align}

\newpage

\subsection{Case `A' - the tilted flat but rounded punch with unequal radii}\label{appendix:caseA}

\hspace{0.4cm}Case `A' considers that the curvature of the middle segment, $1/R$, goes to $0$. As said before, an indenter  like this is often referred to as the flat but rounded punch. The symmetrical case has been studied thoroughly in the literature \cite{Vazquez_2009}, \cite{Ciavarella_1998_2}. Here, asymmetry is introduced to the problem in a sense that the radii can be unequal and the normal load is applied at an offset, $s$. The problem is of high interest to the aerospace industries since this kind of punch can be used to represent a dovetail flank contact. A closed-form solution is found by taking the limit $R \to \infty$ for equation \eqref{pressure_abbreviated} and the respective side conditions, i.e. equations \eqref{consistency_condition}, \eqref{equilibrium_condition_1}, and \eqref{equilibrium_condition_2}.

\begin{figure}[H]
	\centering
	\includegraphics[scale=0.35, trim=100 1870 0 70, clip]{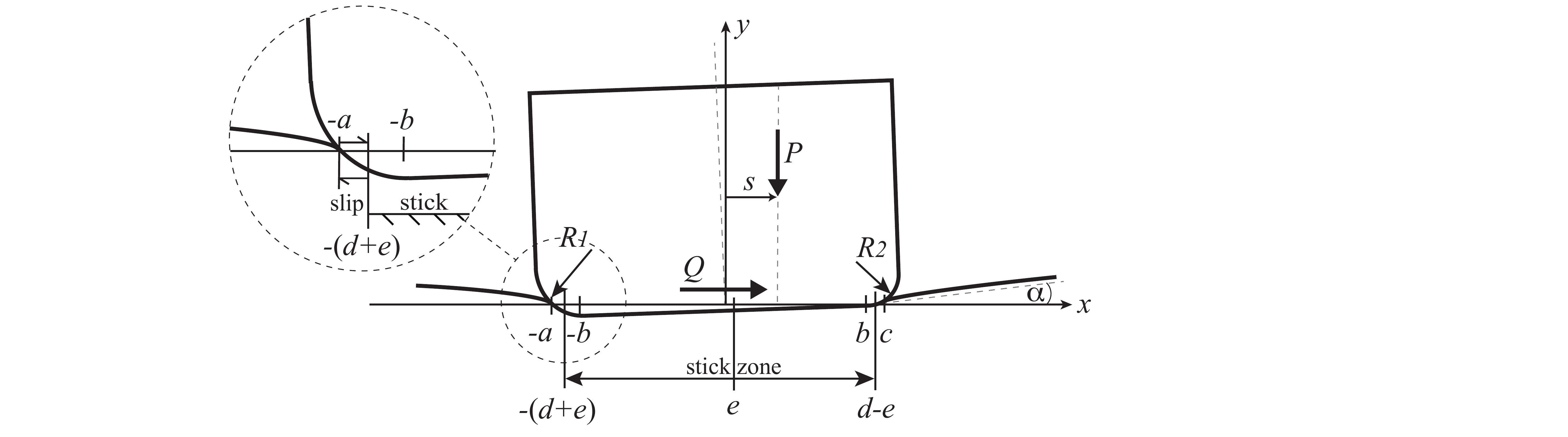}
	\caption{Illustration of the tilted flat but rounded punch with unequal radii}
	\label{fig:illustration_case_C}	
\end{figure}

 It follows that $k_{1} \to 1$ and $k_{2} \to 1$, so that the pressure distribution is given in compact and closed form by
\begin{align}
p\left( x\right) &=-\frac{w(x)}{\uppi A}\left( F_{0} +\frac{\alpha R_{1} + b +x }{R_{1}\, w(x) }\, F_{1}(x) +\frac{\alpha}{w(x)} F_{2}(x)  -\frac{-\alpha R_{2} + b-x }{R_{2}\, w(x)}\, F_{3}(x)
\right)  \text{   ,}
\end{align}
or in full expression as
\begin{align}\label{pressure_distribution_compact_1}
p(x)=&-\frac{\sqrt{(a+x)(c-x)}}{\uppi A}\Bigg( \frac{\arccos\left[\frac{%
		-a+2b+c }{ a+c }\right] }{R_{1}}+\frac{\arccos\left[\frac{ a+2b-c }{ a+c }\right] }{R_{2}}+\Bigg. \nonumber \\
&\left.{}+{} \frac{(b+x)}{R_{1}\sqrt{(a+x)(c-x)}} \ln \left\vert \frac{(a+c)(b+x)}{(c-a)(x-b)+2(ac+bx)+2\sqrt{%
		(a-b)(c+b)(c-x)(a+x)}}\right\vert +\right. \nonumber \\
&\, \, \Bigg. {}+{} \frac{(b-x)}{R_{2}\sqrt{(a+x)(c-x)}}\ln
\left\vert \frac{(a+c)(b-x)}{(c-a)(x+b)+2(ac-bx)+2\sqrt{(a+b)(c-b)(c-x)(a+x)}%
}\right\vert \,  \Bigg)  \text{   .}
\end{align}

The consistency condition is given by
\begin{align}\label{consistency_cond_A}
\alpha =&\frac{1}{\uppi} \left(+ \frac{\sqrt{(a-b) (b+c)}}{R_{1}}+\frac{a-2b-c}{2 R_{1}}\arccos\left[ \frac{-a+2b+c}{ a+c}\right]+ \right.\nonumber \\
&\quad \, \, \left. {}-{}\frac{\sqrt{(a+b) (c-b)}}{R_{2}}+\frac{a+2b-c}{2 R_{2}}\arccos\left[ \frac{a+2b-c}{ a+c}\right] \right) \text{   .}
\end{align}

Imposing normal and rotational equilibrium, the two side conditions are
\begin{align}\label{vertical_equilibrium_A}
\frac{4 PA}{\left( a+c\right) ^{2}}=&  {}-{} \frac{\arccos\left[ \frac{-a+2b+c }{ a+c }\right] }{2R_{1}}-\frac{\arccos\left[ \frac{ a+2b-c }{ a+c }\right] }{2R_{2}} {}+{} \nonumber  \\
& {}+{}\frac{(-a+2b+c)\sqrt{(a-b)(c+b)}}{\left( a+c\right) ^{2}R_{1}}+\frac{(a+2b-c)\sqrt{(a+b)(c-b)}}{\left( a+c\right) ^{2}R_{2}} \text{  ,} 
\end{align} 
and  
\begin{align}
\frac{4 PA s}{\left( a+c\right) ^{2}}=& \left( a-c\right)\left({}+{}  \frac{\arccos\left[ \frac{ -a+2b+c }{ a+c }\right] }{4R_{1}}+\frac{\arccos\left[ \frac{ a+2b-c }{ a+c }\right] }{4R_{2}} {}+{}\right. \nonumber \\ 
&\quad \quad \quad \, \, \, \, \, \left. {}+{} \frac{\left( 
	\frac{a^{2}-2b^{2}+c^{2}}{\left( a-c\right) }+\frac{a-2b-c}{2}\right) \sqrt{%
		(a-b)(c+b)}}{3\left( a+c\right) ^{2}R_{1}}{}-{} \right. \nonumber \\
&\quad \quad \quad \, \, \, \, \, \left.{}-{} \frac{\left( \frac{%
		a^{2}-2b^{2}+c^{2}}{\left( a-c\right) }+\frac{a+2b-c}{2}\right) \sqrt{%
		(a+b)(c-b)}}{3\left( a+c\right) ^{2}R_{2}} {}\,{} \right) \text{  .} 
\end{align}

\subsection{Case `B' - the rounded punch with a concave central segment}\label{appendix:caseB}
\begin{figure}[H]
	\centering
	\includegraphics[scale=0.35, trim=100 1870 0 70, clip]{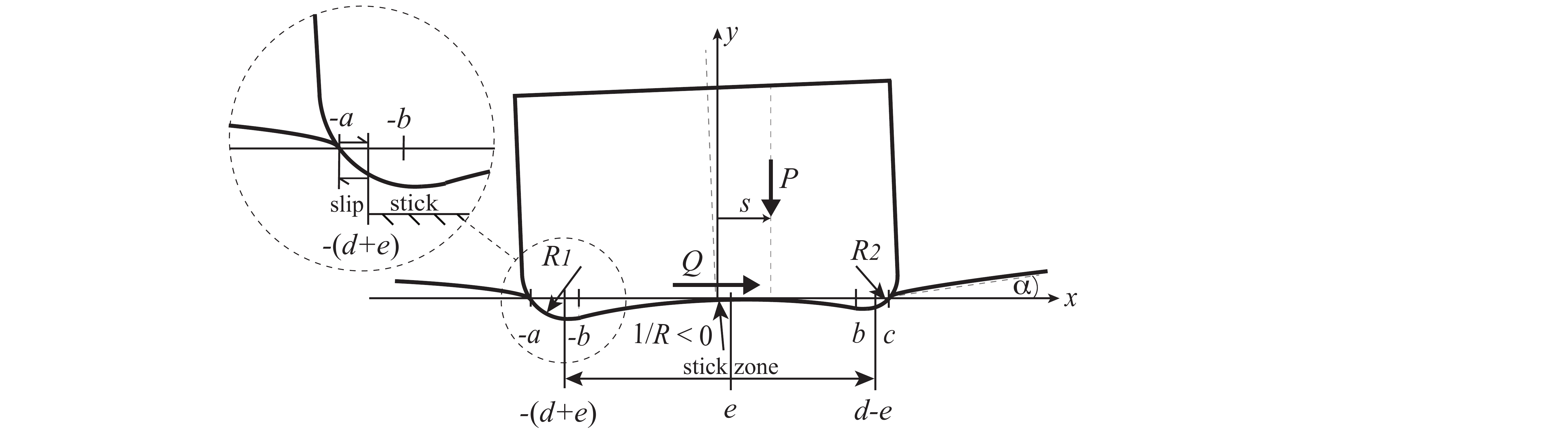}
	\caption{Illustration of the tilted punch comprised of three different rounded segments, where the middle part is concave}
	\label{fig:illustration_case_B}	
\end{figure}

\hspace{0.4cm} Consider a punch comprised of two convex edges of unequal radii and a concave middle portion as depicted in figure \ref{fig:illustration_case_B}. Additionally a moment might be applied in order to introduce tilt. The solution found and presented for the generic three-segment punch still applies here, given the central curvature, $1/R$, is negative. Hence, the equations will not be repeated for brevity. Obviously, a requirement for the solution to be valid is that the contact remains simply connected.

\newpage

\subsection{Case `C' - the tilted punch comprised of two adjoined unequal half-cylinders}\label{appendix:caseC}
\hspace{0.4cm} The last case to be considered is the tilted rounded punch with unequal radii, i.e. the flat portion is removed from the indenter by taking the limit $b \to 0$. 
\begin{figure}[H]
	\centering
	\includegraphics[scale=0.35, trim=100 1870 0 50, clip]{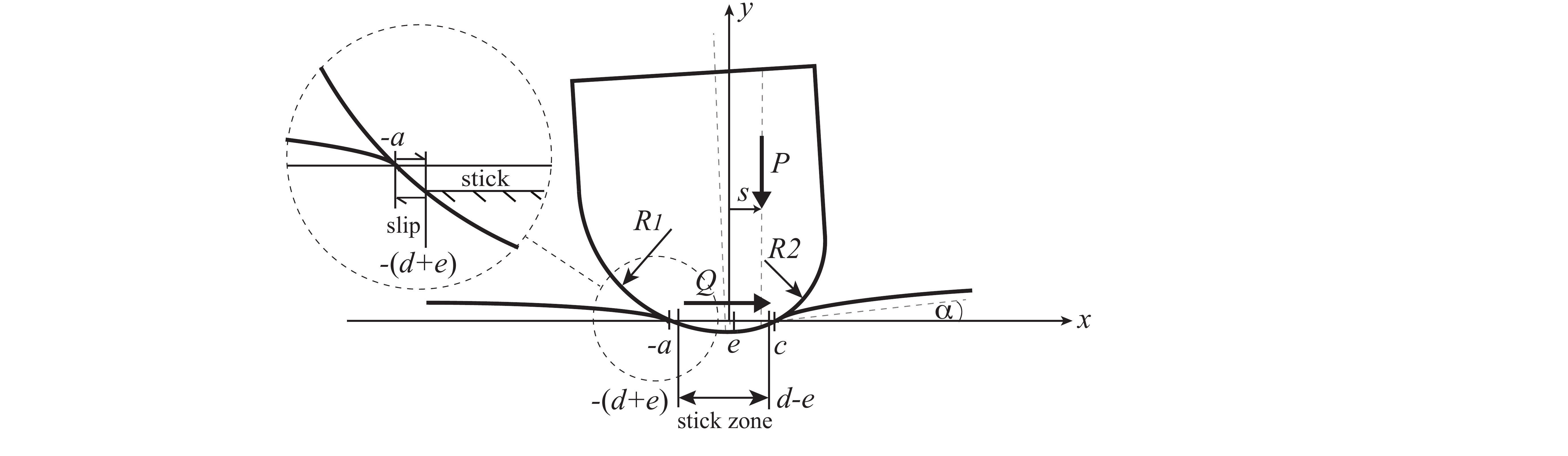}
	\caption{Illustration of the tilted rounded punch with unequal radii}
	\label{fig:illustration_case_D}	
\end{figure}

The pressure distribution, $p(x)$, simplifies to the following expression 
\begin{align}\label{pressure_distribution_compact_2}
p(x)=& -\frac{\sqrt{(a+x)(c-x)}}{\uppi A} \left( \Big(\frac{1}{R_1} + \frac{1}{R_2}\Big) \frac{\uppi}{2} +\Big( \frac{1}{R_1} - \frac{1}{R_2} \Big) \arcsin\left[\frac{a-c}{a+c}\right]  {}+{} \right. \nonumber \\
&\left. {}+{} \Big(\frac{1}{R_1}-\frac{1}{R_2}\Big) \frac{x}{\sqrt{(a+x)(c-x)}} \ln \left\vert \frac{x (a+c)}{2 a c+(c-a) x+2 \sqrt{a c (c-x) (a+x)}}\right\vert  \right)  \text{.} 
\end{align}

The consistency condition is given by
\begin{align}\label{consistency_cond_2}
\alpha = & \left(\frac{1}{R_1} - \frac{1}{R_2}\right) \frac{\sqrt{a c}}{\uppi}  + \left(\frac{1}{R_1} - \frac{1}{R_2} \right) \frac{a-c}{2 \uppi}   \arcsin\left[\frac{a-c}{a+c}\right] + \left(\frac{1}{R_1} + \frac{1}{R_2}\right) \frac{a-c}{4} \text{  . }
\end{align}

Imposing normal and rotational equilibrium, the two side conditions are
\begin{align}\label{vertical_equilibrium_2}
\frac{4PA}{(a+c)^2} = & \left(\frac{1}{R_1} - \frac{1}{R_2}\right) \frac{ (c-a) \sqrt{a c} }{ (a+c)^2}  - \frac{1}{2} \left(\frac{1}{R_1} - \frac{1}{R_2}\right)  \arcsin\left[\frac{a-c}{a+c}\right] -  \left(\frac{1}{R_1}+ \frac{1}{R_2}  \right)\frac{\uppi}{4}  \text{   ,}
\end{align} 
and  
\begin{align}
\frac{4PAs}{(a+c)^2}= & \frac{a-c}{12} \left(  {}+{} \left(\frac{1}{R_1} - \frac{1}{R_2}\right) \frac{4 \sqrt{a c} }{(a+c)^2} \left( \frac{a^2 + c^2}{a-c} + \frac{a-c}{2}\right)\right. {}+{}\nonumber \\
& \quad \quad \quad \, \left. {}+{}\left(\frac{1}{R_1} - \frac{1}{R_2}\right) 3  \arcsin\left[\frac{a-c}{a+c}\right] + \left(\frac{1}{R_1}+ \frac{1}{R_2} \right)  \frac{ 3 \uppi}{2}\, \right) \text{   .}
\end{align}

\end{appendix}

\end{document}